# Hybrid integrated narrow linewidth laser with external distributed optical feedback from a silicon strip waveguide

*Da Wei, Leilei Shi\*, Yujia Li, Minzhi Xu, Chaoze Zhang, Xianming Huang, Jianxian Yu, Lei Zhai, Wenxuan Huang, Huan Tian, And Tao Zhu\**

Key Laboratory of Optoelectronic Technology & Systems (Education Ministry of China), Chongqing University, Chongqing 400044, China

KEYWORDS: narrow linewidth laser, hybrid integrated, distributed feedback, surface radiation, waveguide

**ABSTRACT:** External optical feedback via Rayleigh scattering from an integrated microresonator or an optical fiber has been demonstrated to significantly narrow the intrinsic linewidth of semiconductor lasers. Wavelength matching between the lasing cavity and the external high-Q microresonator is required to accumulate Rayleigh scattering based optical feedback. Optical fiber can provide Rayleigh scattering based optical feedback for any lasing wavelength. However, optical fibers hundreds of meters or even kilometers long are required for the accumulation of Rayleigh scattering based optical feedback, hindering the integration of narrow linewidth lasers. Here, we present an integrated scheme that collects distributed feedback signal with weak wavelength dependence by exploiting surface radiation in a silicon strip waveguide. The effects of waveguide width on the intensities of the surface radiation and distributed optical feedback signal are first numerically analyzed by introducing a collection coefficient. Numerical calculations show that a 1 μm-wide strip waveguide yields optimal performance for excitation and collection of distributed optical feedback, which is also experimentally verified by measuring the feedback signal with an optical frequency-domain reflectometry. Benefitting from the enhanced distributed optical feedback that is 34.72 dB higher than that in a single-mode fiber, the hybrid integrated laser demonstrates an intrinsic linewidth of 1.52 kHz, a side-mode suppression ratio (SMSR) of 74.71 dB, and a frequency noise of 24.44 Hz²/Hz. Furthermore, within a maximum allowable wavelength tuning range of 2.342 nm, the linewidth narrowing ratio depends little on the wavelength for all the waveguides with different widths. The hybrid integrated narrow linewidth laser has significant potential for precision measurement and communication. Additionally, the method for efficiently collecting distributed feedback signal provides a feasible solution for integrating random lasers and chaotic lasers.

■ INTRODUCTION

Miniaturized narrow linewidth lasers are highly demanded for various coherent applications with stringent volume and weight constraints, such as gyroscope, [1, 2] LiDAR, [3, 4] spectroscopy, [5] optical frequency synthesis, [6, 7] and microwave photonics. [8-10] III–V semiconductor lasers offer unparalleled advantages in size, weight, power consumption, and cost compared to fiber and solid-state lasers. However, the linewidth of a laser diode, such as a distributed feedback (DFB) laser, is usually on the order of hundreds of kilohertz to megahertz. Importing external optical feedback has been widely used to significantly narrow the linewidth and reduce frequency noise of a semiconductor laser. Hybrid integration of the lasing cavity with an external cavity offers high performance and design flexibility. Integrated on-chip waveguides, including Bragg gratings [11, 12] and integrated add-drop microcavities, [13-17] have been implemented to narrow the linewidth to kilohertz or less. However, additional control mechanisms, such as electro-optic or thermo-optic negative feedback, must be adopted to meet the strict wavelength-matching requirement, which complicates the system configuration to some extent.[18-21] Especially for arrayed lasers, increasing the number of gratings for wavelength matching increases the difficulty of aligning the lasing cavity with the external passive cavity.

In addition to grating or microcavity based single-point feedback, Rayleigh scattering based distributed weak feedback from high-$Q$ microcavities and optical fibers has also been widely used to narrow the linewidth and reduce frequency noise of semiconductor lasers. [22-31] However, the inherently weak Rayleigh scattering requires ultra-high-Q microcavities or ultra-long optical fibers to accumulate the distributed feedback signal, posing significant fabrication challenges or hindering the integration of narrow linewidth semiconductor lasers.

In this context, we utilize surface radiation combined with internal Rayleigh scattering and leverage the high refractive-index contrast of the silicon-on-insulator (SOI) platform to significantly improve distributed feedback signal excitation and collection. Experimental results show that the intensity of the distributed feedback signal from a 1 μm width scattering waveguide (SWG) is ~37 dB higher than that of a standard single-mode fiber. The linewidth of a DFB laser diode is narrowed to 1.52 kHz by the distributed feedback from the waveguide hybrid integrated with the lasing cavity. The side-mode suppression ratio (SMSR) and the white frequency noise of the hybrid integrated laser are 74.71 dB and 24.44 $Hz^2/Hz$, respectively. The weak wavelength-dependence of the distributed feedback is also experimentally verified by the nearly unchanged linewidth over a continuous wavelength tuning range of 2.342 nm. The work demonstrated here provides a method for an integrated on-chip waveguide that collects the distributed feedback signal without wavelength selectivity. Such a waveguide design methodology can also be extended to other platforms, such as silicon nitride and lithium niobate, paving the way for the development of integrated narrow linewidth lasers, especially wavelength-tunable narrow linewidth lasers.

■ **PRINCIPLE AND EXPERIMENTAL SETUP**

The schematic diagram of the proposed hybrid narrow linewidth laser is shown in Figure 1. Light from the DFB laser diode is injected into a 1×2 multimode interferometer (MMI) via edge coupling, where it is split into two paths. One propagates along the SWG after passing through a mode converter, in which the distributed feedback signal is excited and fed back into the DFB lasing cavity. The other is coupled to a lens fiber array (FA) and then characterized by a delayed self-heterodyne interferometry (DSHI). The DSHI consists of two 3 dB optical couplers (OC), a polarization controller (PC), a delay fiber, an acousto-optic modulator (AOM) with a frequency shift of 100 MHz, a photo detector (PD, Thorlabs, PDB 450C), and an electrical spectrum analyzer (ESA, FSV 3030, ROHDE & SCHWARZ). As shown in the inset of Figure 1, the DFB laser diode and the external cavity are hybrid-integrated and packaged within a standard 14-pin butterfly housing. A thermoelectric cooler (TEC) is integrated beneath the DFB laser chip to enable independent thermal control.

The analytical model of waveguide surface radiation is shown in Figure 2, where a Cartesian coordinate system is established with its origin at the center of the waveguide's cross-section. The waveguide is defined by its semi-width $d$ (along the $x$-direction), height $h$ (along the $y$-direction), and length $l$ (along the $z$-direction). The refractive indices of the waveguide and the cladding layer are $n_1$ and $n_2$, respectively. Unlike the cylindrical optical fiber with an ideally smooth core-cladding interface independent of the azimuthal angle, only the top and bottom interfaces between the waveguide and the cladding are assumed to be ideally smooth since the optical field is primarily confined to a quasi-two-dimensional space. The roughness on the left and right waveguide-cladding surfaces of the photo-lithographed waveguide results in more scattering signal.

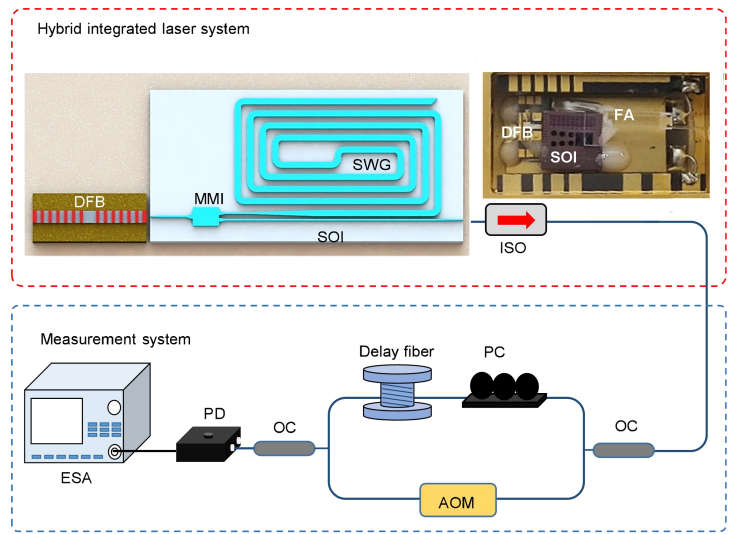

Figure 1. Schematic diagram of the hybrid integrated narrow linewidth laser with external distributed feedback and the short DSHI-based measurement system.

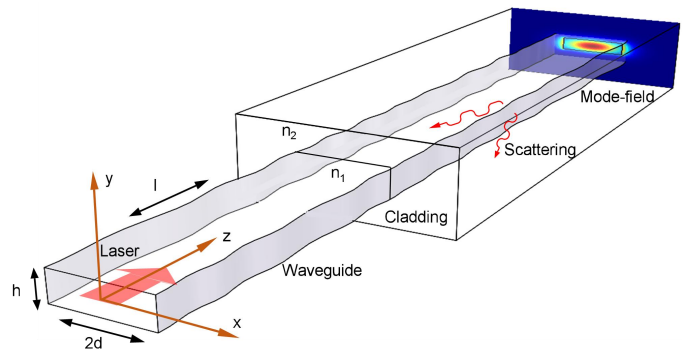

Figure 2. Model of waveguide surface radiation

According to the equivalent current method, the scattering coefficient $\alpha_b$ induced by rough waveguide boundaries can be written as : [32]

$$\alpha_r = \varphi^2(d)(n_1^2 - n_2^2)^2 \frac{k_0^3}{4\pi n_1} \int_0^\pi \tilde{R}(\beta - n_2 k_0 \cos\theta) d\theta \qquad (1)$$

where $\varphi(d)$ is the amplitude of the modal field at the waveguide surface, $\tilde{R}(\Omega) = \int_{-\infty}^{\infty} R(u) \exp(i\Omega u) du$ is the spectral density function of the surface roughness, $k_0 = 2\pi/\lambda$ is the free-space wavenumber, $\beta = k_0 n_{\text{eff}}$ is the propagation constant, and $\theta$ is the azimuthal angle ranging from 0 to π. For simplicity, we define six dimensionless parameters: the normalized refractive index contrast $\Delta = (n_1^2 - n_2^2)/2n_1^2$, the normalized frequency $V = k_0 d \sqrt{2n_1^2 \Delta}$, the normalized waveguide parameter $U = d\sqrt{n_1^2 k_0^2 - \beta^2}$, the attenuation parameter $W = d\sqrt{\beta^2 - n_2^2 k_0^2}$, the normalized correlation length $\chi = W * L_c / d$ with $L_c$ the correlation length of the surface roughness, and the waveguide confinement strength indicator $\gamma = n_2 V / (n_1 W \sqrt{\Delta})$. By applying the half-angle substitution $t = \tan(\theta/2)$, the integral can be evaluated exactly as

$$\alpha_r = \frac{\varepsilon^2}{\sqrt{2} k_0 d^4 n_1} g(V) f_e(\chi, \gamma) \qquad (2)$$

where $g(V) = U^2 V^2/(1+W)$ is the geometric factor, and $\varepsilon^2 = R(0)$ is the surface roughness, $f_e(\chi,\gamma)$ depends $\chi$ and $\gamma$ as

$$f_e(\chi,\gamma) = \frac{\chi \left\{ \left[(1+\chi^2)^2 + 2\chi^2\gamma^2\right]^{1/2} + 1 - \chi^2 \right\}^{1/2}}{\left[(1+\chi^2)^2 + 2\chi^2\gamma^2\right]^{1/2}} \qquad (3)$$

The upper limit of the loss coefficient is determined by the maximum of $f_e(\chi,\gamma)$. For the waveguide with strong optical field confinement ($\gamma \approx 1$), $f_e(\chi,\gamma) \approx 1$, which results in that

$$\alpha_r \leq \frac{\varepsilon^2}{2k_0 d^4 n_1} g(V) \tag{4}$$

In addition to surface radiation, other losses, $\xi$ originating from Rayleigh scattering, and $\zeta$ originating from material absorption, also contribute to the total loss. Thus, the total loss coefficient is:

$$\alpha_{loss} = \alpha_r + \xi + \zeta \tag{5}$$

The excitation coefficient of total scattering can be expressed as:

$$\alpha_s = \alpha_r + \xi \tag{6}$$

A portion of the scattered light is captured by the waveguide and propagates in the backward direction. Given that silicon waveguides exhibit strong optical field confinement due to the high refractive index contrast, in this case, the backscattering collection coefficient $S$ can be written as: [33]

$$S = \frac{3\lambda^2}{8\pi^2 n_{eff}^2 w_x w_y} \cdot \gamma \cdot F \cdot G \tag{7}$$

where $w_x$ and $w_y$ are the mode field radii in the $x$ and $y$ directions, respectively, $F = 1 + \frac{2}{3}\Delta + \frac{1}{2}\Delta^2$ is the strong guidance correction factor, and $G = (3 + 3\langle\cos^2\theta\rangle)/4$ is the polarization correction factor. For strongly guiding silicon waveguides ($\Delta \approx 0.414$), the optical field is almost entirely confined within the waveguide, so $\gamma \approx 1$. For the mode $TE_{10}$, $G \approx 1$, and $S$ can be simplified into

$$S = \frac{3\lambda^2}{8\pi^2 n_{eff}^2 w_x w_y} \cdot F \quad (8)$$

The backscattered power can then be expressed as: [33]

$$P_s(z) = P_0 \cdot \exp(-2\alpha_{loss} z) \cdot \alpha_s dz \cdot S \quad (9)$$

where $P_0$ is the incident power.

We numerically simulate the excitation coefficient of scattering $\alpha_s$, the total loss coefficient $\alpha_{loss}$, the collection coefficient $S$, and the backscattered power $P_s$ for a silicon waveguide with a $SiO_2$ cladding using the parameters listed in Table 1. Figure 3(a) shows the calculated $\alpha_s$ as a function of waveguide width (2d), indicating that as the width increases, $\alpha_s$ decreases gradually and saturates beyond 1 μm. The dependence of the effective refractive index $n_{eff}$ is shown in Figure 3(b), from which it can be seen that $n_{eff}$ tends to saturate as the waveguide width exceeds 1 μm. The scattering collection coefficient $S$ as a function of increasing waveguide width is shown in Figure 3(c). According to Eq. (7), $S$ is primarily determined by $n_{eff}$, $w_x$, $w_y$, and $\Delta$ collectively. An increase in waveguide width enlarges the mode field area, thereby indirectly regulating $S$.

The power $P_s$ at $z = 0$ in Figure 3(d) shows that $P_s$ decreases as the waveguide width increases. The distribution of $P_s$ along the z-direction, as shown in Figure 3(e), indicates that scattering is predominantly localized near the input when $d$ is small, resulting in reduced overall feedback efficiency and a more concentrated phase distribution. When the waveguide is wider than 1 μm, $\alpha_s$ no longer decreases, allowing the feedback to distribute over a longer waveguide and improving the overall feedback efficiency. However, $P_s$ will continue to decrease as $S$ decreases. Therefore, the 1 μm-wide waveguide is an optimal choice for balancing scattering loss and collection efficiency.

Table 1. Parameters used for the numerical simulation

| Parameter | Value | | | Parameter | | Value | |
|---|---|---|---|---|---|---|---|
| $\lambda$ (μm) | 1.55 | | | $\sigma$ (nm) | | 1.6 | |
| $n_1$ | 3.47 | | | $Lc$ (nm) | | 50 | |
| $n_2$ | 1.44 | | | $h$ (nm) | | 220 | |
| Parameter | Value | | | | | | |
| $2d$ (μm) | 0.45 | 0.65 | 0.86 | 1 | | 1.5 | 2 |
| $n_{eff}$ | 2.35 | 2.60 | 2.68 | 2.74 | | 2.80 | 2.82 |
| $w_x$ (μm) | 0.22 | 0.27 | 0.33 | 0.37 | | 0.52 | 0.68 |
| $w_y$ (μm) | 0.21 | 0.21 | 0.21 | 0.21 | | 0.21 | 0.21 |

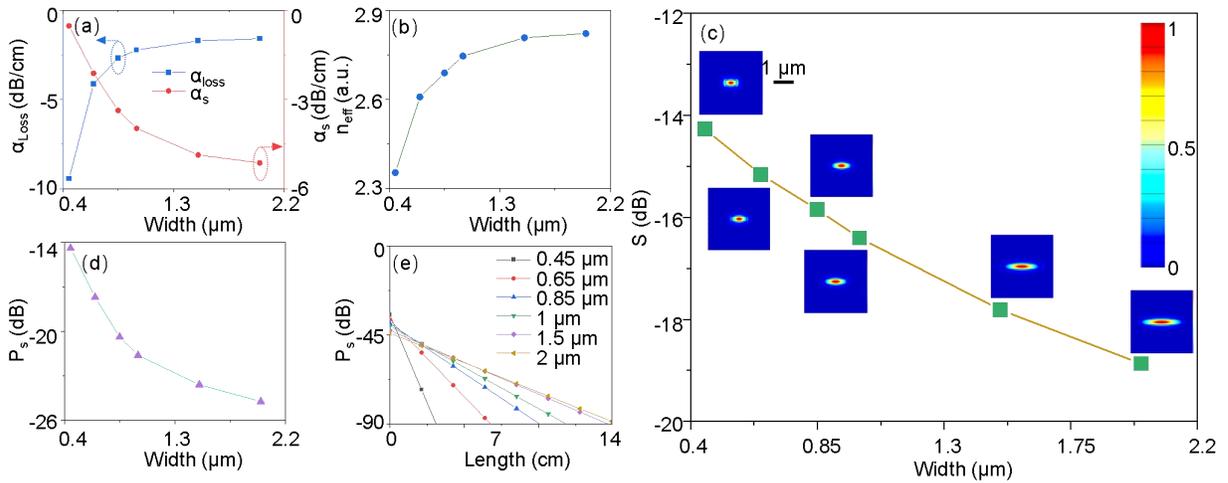

Figure 3. Numerical calculation of (a) $\alpha_s$, and $\alpha_{loss}$ (b) $n_{eff}$, (c) $S$, and (d) $P_s(z=0)$ with respect to the waveguide width. (e) Dependence of the power $P_s$ along the z-axis on the waveguide width.

In addition to the SWG, the 50 μm-long tapered mode field converter, with its two ends connected to the 450 nm-wide single-mode waveguide and the SWG, respectively, also affects the loss. To study light propagation in more detail, we conduct a simulation using the finite-difference time-domain (FDTD) method. The parameters, except for the straight length of the SWG, are identical to those of the fabricated device. Figures 4(a) to 4(f) show the simulated mode field for the waveguide with a width of 0.45 μm, 0.65 μm, 0.85 μm, 1 μm, 1.5 μm, and 2 μm, respectively. The SWG cannot strictly confine the optical field due to the higher-order modes excited by widening the waveguide. The deviation becomes particularly noticeable when the waveguide width exceeds 1 μm. The insets of Figs. 4(a) to (f) show the mode field distribution on the waveguide cross-section before the mode field converter. The losses of the mode field converter under different waveguide widths are then statistically analyzed, as shown in Figure 4(g). The loss increases more rapidly as the width exceeds 1 μm.

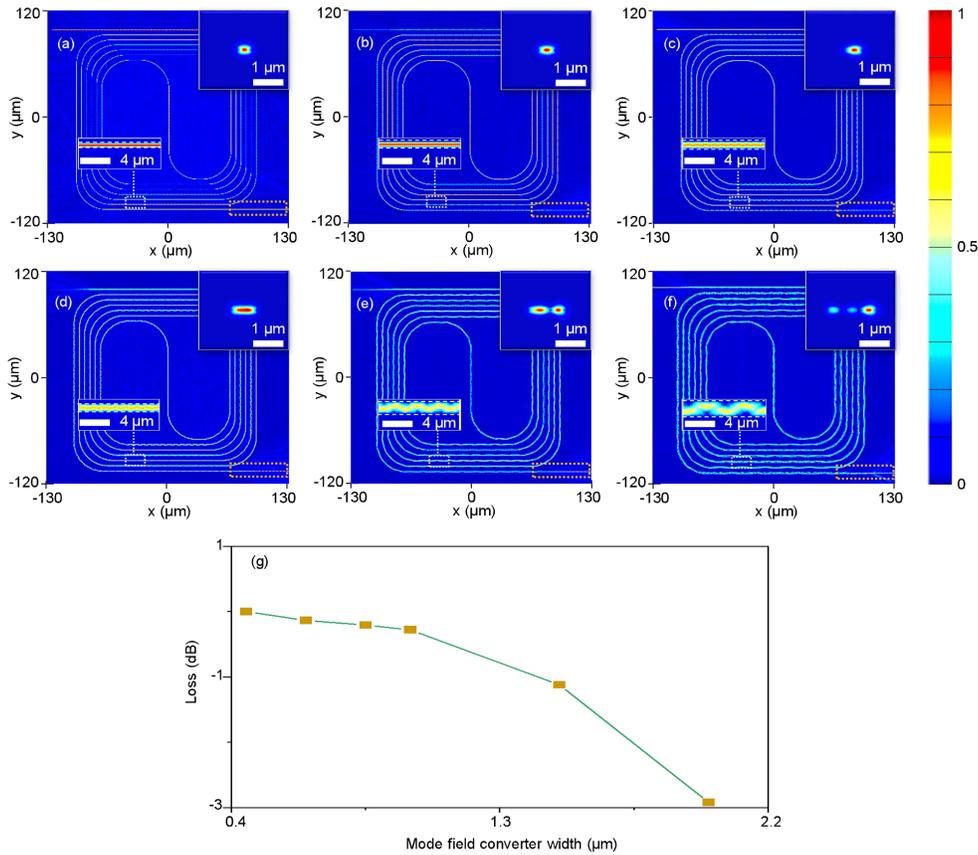

Figure 4. Optical field distributions for the waveguide with a width of (a) 0.45 μm, (b) 0.65 μm, (c) 0.85 μm, (d) 1 μm, (e) 1.5 μm, and (f) 2 μm, in which the inset shows the mode field

distribution on the waveguide cross-section. (g) Dependence of the loss on the mode field converter width.

■ **EXPERIMENTAL RESULTS AND DISCUSSION**

We fabricated SWGs on a SOI platform with a silicon layer thickness of 220 nm and widths varying from 0.45 μm, 0.65 μm, 0.85 μm, 1 μm, to 2 μm. Edge couplers with a mode field diameter of 3.5 μm were fabricated at both ends of the waveguide. Figure 5(a) shows the photograph of the packaged SOI chip with an FA. An optical frequency-domain reflectometry (OFDR) system is then used to characterize the backscattering signal from the packaged SOI waveguide, [34, 35] as shown in Figure 5(b). The inset of Figure 5(b) shows the micrograph of the SWG with a length of ~14 cm. Figures 5(c)-(g) show the distributions of the backscattering signal along the waveguide with different widths. It can be seen that a wider waveguide results in a longer effective backscattering length and higher scattering power, which is consistent with the theoretical analysis. The backscattering intensity of the 1 μm-wide waveguide is 34.72 dB higher than that of a standard single-mode fiber. Figure 5(h) shows the loss coefficient $α_{loss}$ obtained from the data within the dashed rectangular areas in Figure 5(c)-(g), which agrees well with the numerical simulation. When the waveguide width increases from 1 μm to 2 μm, the scattering coefficient decreases only slightly; however, the overall backscattering intensity is significantly reduced. The reason is that the backscattering collection coefficient $S$ decreases with increasing width, and a wider waveguide is more prone to excite higher-order modes, which introduce additional losses through the mode field converter.

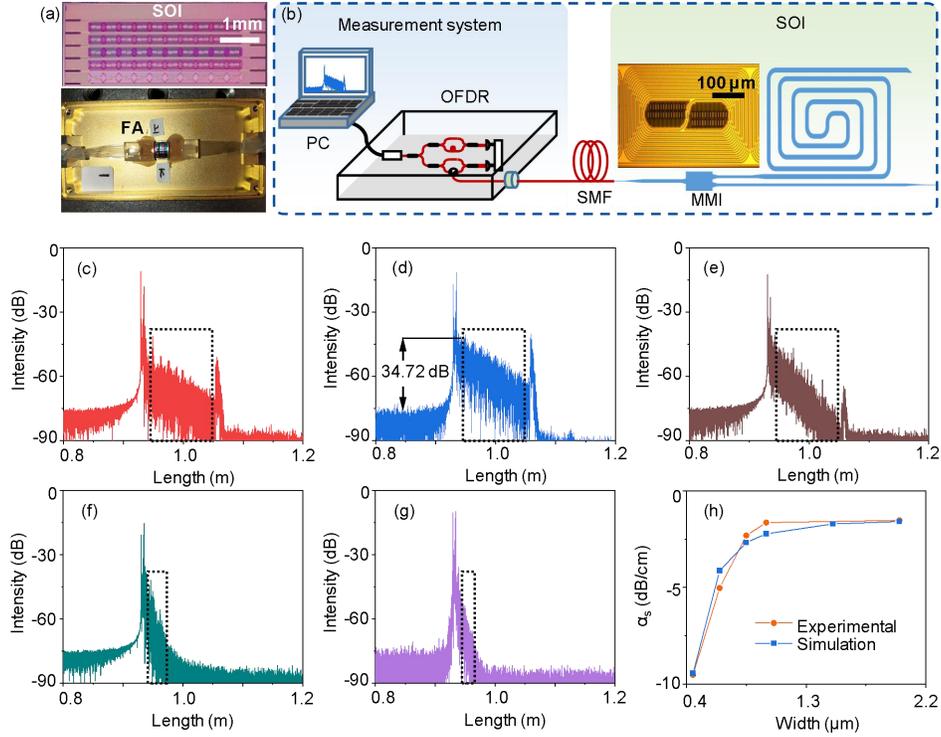

Figure 5. (a) Photograph of the packaged SOI chip with the FA. (b) Backscattering characterization based on an OFDR. Backscattering intensity distribution along the waveguide with a width of (c) 2 μm, (d) 1 μm, (e) 0.85 μm, (f) 0.65 μm, and (g) 0.45 μm. (h) Dependences of the measured and simulated backscattering attenuation coefficients on the waveguide width.

Then, distributed feedback from SWG with different widths is successively injected into the same DFB laser diode to study the effect of SWG width on linewidth narrowing and frequency noise reduction. The beat spectra measured by the DSHI with a 50-km delay fiber are shown in Figure 6(a). A DFB laser with optical feedback demonstrates a narrower linewidth than the free-running DFB laser, indicating the capability of distributed feedback for linewidth narrowing. Moreover, the linewidth narrowing ratio increases monotonically as the SWG width is less than 1 μm, then decreases due to a lower overall backscattering collection efficiency. Thus, distributed feedback from the SWG with a width of 1 μm yields the narrowest linewidth in our current experiments. It should be noted that a narrower linewidth could be expected if a technique for fabricating longer waveguides with higher scattering excitation efficiency becomes available in the future. Meanwhile, the frequency noise of the DFB with external distributed feedback from SWG with different widths is also measured by replacing the 50-km delay fiber in the DSHI with a 10-m optical fiber, as shown in Figure 6(b). As expected, the DFB laser diode with distributed feedback from the 1 μm-wide SWG exhibits the lowest white frequency noise.

For a clearer comparison, the beat spectra and frequency noise spectra of the DFB laser diode without and with optical feedback from the SWG with a width of 1 μm are replotted in Figure 7. The free-running linewidth of the DFB laser was 125.23 kHz. When distributed feedback from the SWG with a width of 1 μm is injected into the lasing cavity, the linewidth narrows from 125.23 kHz to 1.52 kHz, as shown in Figure 7(a). The side-mode suppression ratio (SMSR) is improved to 74.71 dB, which is 32.91 dB higher than that of the free-running DFB laser diode, as shown in Figure 7(b). The frequency noise spectra are shown in Figure 7 (c). The frequency noise of the laser with feedback from the 1 μm-wide SWG is 24.44 Hz²/Hz at the offset frequency of 1 MHz, which is 38.93 dB lower than that of the free-running DFB laser.

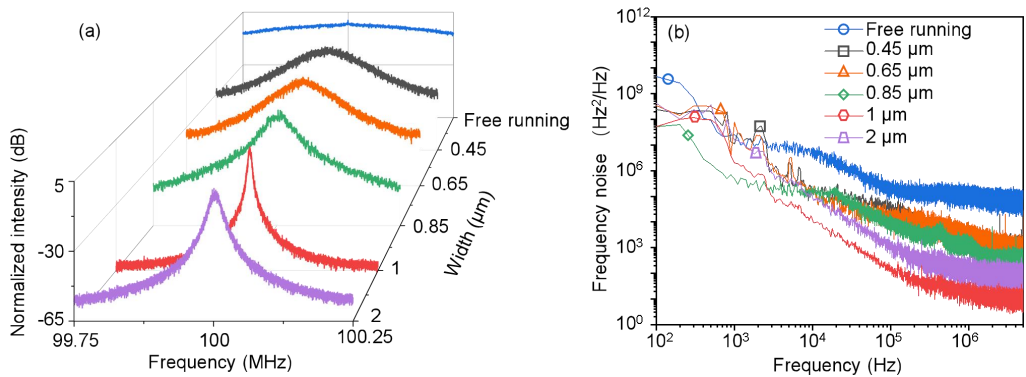

Figure 6. (a) Linewidth and (b) Frequency noise of the DFB laser diode without (free-running) and with distributed feedback from SWG with different widths.

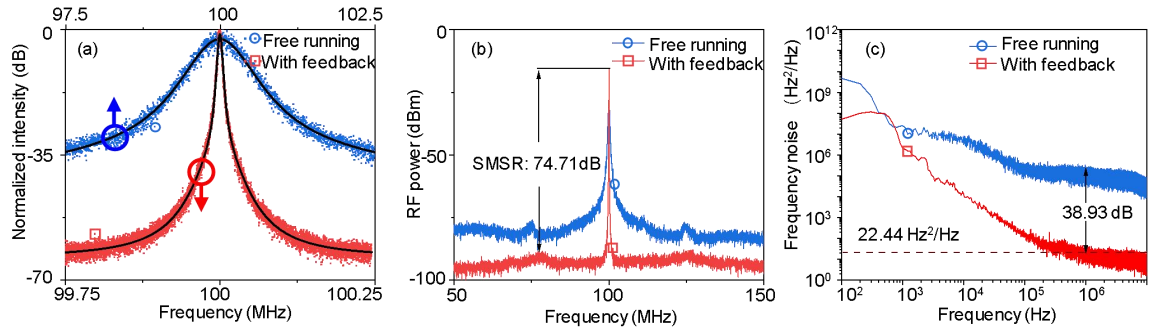

Figure 7. (a) Linewidth, (b) SMSR, and (c) Frequency noise of the DFB laser diode without and with optical feedback from the SWG with a width of 1 μm.

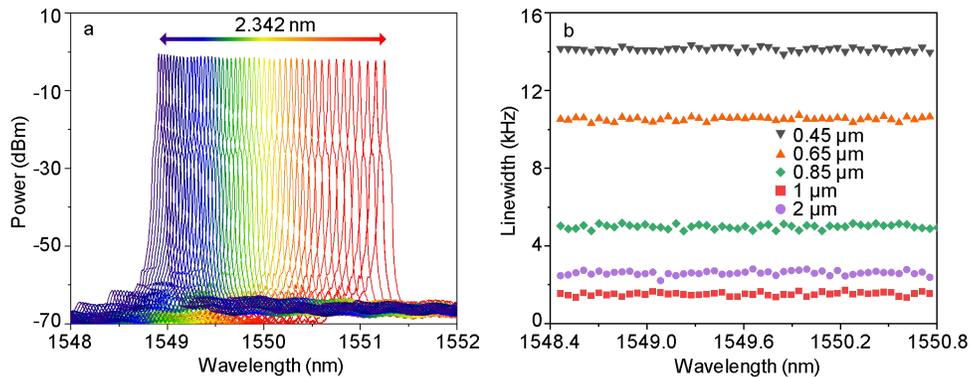

Figure 8. (a) Optical spectrum and (b) linewidth of the DFB laser diode as the wavelength tunes from 1548.411 nm to 1550.753 nm.

Finally, we measured the linewidth of the DFB laser with distributed feedback from the SWGs with different widths as the lasing wavelength was tuned from 1548.411 nm to 1550.753 nm by adjusting the thermal resistance of the driver, as shown in Figure 8(a). Figure 8(b) shows the measured linewidth of the DFB laser diode with distributed feedback from SWGs with different widths. Independent of the SWG width, the laser linewidth remained within the same order of magnitude in a wavelength tuning range of 2.342 nm, experimentally verifying the weak wavelength dependence of the distributed feedback from the SWG. Moreover, the linewidth of the DFB laser diode with a 1 μm width SWG is always the narrowest in the whole wavelength tuning range. Therefore, the SWG with wavelength self-adaptive distributed feedback provides an alternative method for integrating a tunable narrow linewidth laser, with advantages in system complexity and control over tunable lasers based on the Vernier effect.

■ **CONCLUSION**

In summary, we have demonstrated a hybrid integrated narrow linewidth laser with surface radiation based external optical feedback. Numerical and experimental studies show that both the waveguide width and the modal conversion loss affect the intensity of the distributed feedback signal in the SWG. An enhanced distributed feedback signal, 34.72 dB above that in a single-mode fiber, has been achieved in a silicon strip waveguide with a 1 μm width. A hybrid-integrated laser combining a III-V DFB laser chip with an optimal silicon strip waveguide demonstrated an intrinsic linewidth of 1.52 kHz and a white frequency noise of 24.44 Hz²/Hz, which is 38.93 dB lower than that of the free-running laser. The nearly constant intrinsic linewidth of the hybrid integrated laser in a maximum allowable tuning range of 2.342 nm experimentally verifies the weak wavelength dependence of the distributed feedback. The method proposed in this work enables the collection of distributed weak feedback signals from chip-scale devices, paving the way for an integrated wavelength-tunable narrow linewidth laser. In addition, the feedback scheme with weak wavelength dependence also holds significant potential for the integrated random lasers and chaotic lasers.

ASSOCIATED CONTENT

**Data Availability Statement**

All data required to reach the conclusion of this study are presented in the manuscript.

AUTHOR INFORMATION

**Corresponding Author**

**Leilei Shi** - Key Laboratory of Optoelectronic Technology & Systems (Education Ministry of China), Chongqing University, Chongqing 400044, China；Email：shileilei@cqu.edu.cn

**Tao Zhu** - Key Laboratory of Optoelectronic Technology & Systems (Education Ministry of China), Chongqing University, Chongqing 400044, China；Email：zhutao@cqu.edu.cn

**Author Contributions**

The manuscript was written through contributions of all authors. All authors have given approval to the final version of the manuscript.

**Funding Sources**

This research was funded by National Natural Science Foundation of China (U23A20378, 62275034)